\documentclass[noshowpacs,amsmath,twocolumn,
aps,prb]
{revtex4-1}
\usepackage{setspace}
\usepackage{amsmath}
\usepackage{graphicx}
\usepackage[nearskip,margin = 0pt]{subfig}
\usepackage{subfig}

\usepackage{verbatim}
\usepackage{amsfonts}
\usepackage{amssymb}
\usepackage{epstopdf} 
\usepackage{xcolor}
\usepackage{color}
\DeclareGraphicsExtensions{.pdf,.eps,.png,.jpg,.mps} 
\usepackage{ragged2e}
\usepackage{hyperref}
\hypersetup{
    colorlinks=true,
    linkcolor=blue,
    citecolor=blue,
    filecolor=blue, 
    urlcolor=blue,
}
\usepackage{mathrsfs}

\usepackage{threeparttable}

\begin{document}

\title{
Photonic timing-engineered solitons for dual-microcomb metrology}
\author{Zihao Wang,$^{1}$ Guoliang Li,$^{1}$ Changrui Liu,$^{1}$ Yue Hu,$^{2,3}$ Zeying Zhong,$^{2,3}$ Changxi Yang,$^{1}$ Junqiu Liu,$^{2,4}$ and Chengying Bao$^{1,*}$ \\
$^1$State Key Laboratory of Precision Measurement Technology and Instruments, Department of Precision Instruments, Tsinghua University, Beijing 100084, China.\\
$^2$International Quantum Academy, Shenzhen 518048, China.\\
$^3$Shenzhen Institute for Quantum Science and Engineering, Southern University of Science and Technology, Shenzhen 518055, China.\\
$^4$Hefei National Laboratory, University of Science and Technology of China, Hefei 230088, China.\\
Corresponding author:  $^*$cbao@tsinghua.edu.cn 
}

\maketitle
\newcommand{\ts}{\textsuperscript}
\newcommand{\tsb}{\textsubscript}


{\bf Dissipative microcavity solitons offer a route to integrate comb-based metrology systems on photonic chips. However, integrated solitons generally lack agile control of comb parameters, particularly pulse timing control, hindering their application in quantum-limited metrology. Here we introduce dynamical soliton trapping to enable optically timing-engineered microcombs (OTEM). By injecting an auxiliary laser to anchor one of the microcomb lines, we create a potential well to trap and steer the soliton. Thus, soliton timing can be engineered by phase modulating the injected laser. Theory and measurement reveal the fast response bandwidth of the OTEM, which enabled a soliton slew rate of 31.3 ps/$\mu$s, surpassing existing time-programmable fiber laser combs by more than two orders of magnitude.
Leveraging the timing scan, we used a single OTEM for single-pixel and parallel ranging by retrieving the phase of the multi-heterodyne beat spectrum. Picometer-scale ranging precision was achieved, establishing new record for optical absolute ranging. Our work can transform timing-engineering of microcavity solitons and endow integrated dual-microcomb metrology systems with enhanced precision. 
}

Random fluctuations of pulse timing or position in a pulse train impact the stability of the corresponding frequency combs \cite{Haus2002noise,Cundiff_OE2008quantum,Matsko_OE2013Jitter,Vahala_NP2021}. On the other hand, recent work has shown active control of pulse timing can enhance the capability of fiber-based comb systems and push the dual-comb measurement precision towards the quantum limit \cite{Newbury_Nature2022time}. Such time-programmable frequency combs (TPFCs) have greatly enhanced the metrology performance in ranging \cite{Newbury_Nature2022time}, spectroscopy \cite{Newbury_NP2024free} and clock networking \cite{Newbury_Nature2023quantum}. Pulse timing-engineering in TPFCs requires a self-referenced comb, which still represents a significant challenge for chip-based frequency microcombs \cite{Kippenberg_Science2018Review,Bowers_NP2022integrated,Diddams_Science2020optical}. Moreover, TPFCs rely upon feedback loops and electronic devices, and have a limited pulse motion speed (i.e., slew rate). Hence, it calls for new physical mechanisms for timing-engineering of microcombs, ideally enabling simpler architectures toward integration alongside faster modulation.

\begin{figure*}[t!]
\begin{centering}
\includegraphics[width=0.96\linewidth]{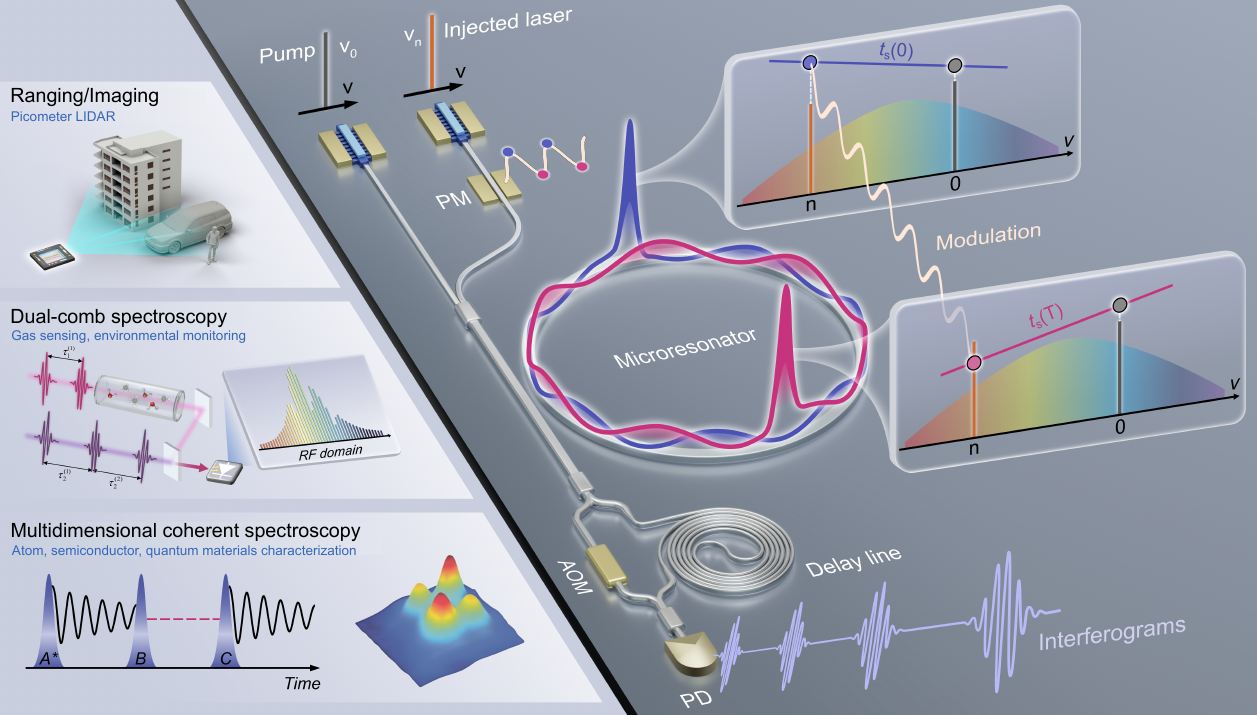}
\captionsetup{singlelinecheck=off, justification = RaggedRight}
\caption{\textbf{Dynamical soliton trapping and optically timing-engineered microcombs (OTEM).} A microcomb line at $\nu_n$ can be injection locked by a control laser. The locked line serves as a trapping well for the microcavity soliton. Phase modulation of the injected laser enables fast steering of soliton position or timing, $t_s(T)$, via dynamical trapping. By delaying an arm of the solitons, dual-comb metrology can be implemented using a single OTEM. The OTEM can be used for ranging/imaging and spectroscopy. PM, phase modulator; PD, photodetector; AOM, acousto-optical modulator. 
} 
\label{Fig1}
\end{centering}
\end{figure*}

In this work, we introduce dynamical trapping of microcavity solitons to enable timing-engineering of microcombs. Dissipative soliton mode-locking in coherently pumped microcavities is refreshing our understanding of soliton physics \cite{Kippenberg_NP2014,Kippenberg_Science2018Review,Weiner_PRL2016,Bowers_Nature2020integrated,Kippenberg_NP2019dynamics,Diddams_NP2017soliton,Bao_PRX2025rhythmic,Vahala_NP2021,Vahala_NP2017Counter}, as well as enabling photonic chip-based spectrometer \cite{Vahala_Science2016,Lipson_SA2018,Vahala_NC2021architecture}, optical frequency synthesizer \cite{Diddams_Nature2017Synthesis}, optical clock \cite{Vahala_Optica2019ACES} and LIDAR \cite{Kippenberg_Science2018Range,Vahala_Science2018Range,Bao_NC2025nanometric}. Solitons can be trapped by a refractive index well created by another field \cite{Menyuk_OL1987,Cundiff_PRL1999,Vahala_Optica2022efficiency}. A special example of soliton trapping in microcavities is that solitons can be trapped by dispersive waves radiated by spatial mode-interactions \cite{Coen_Optica2017,Vahala_NP2021}, which can lead to the formation of soliton crystals \cite{Kippenberg_NP2019dynamics,Diddams_NP2017soliton}. As opposed to using mode-interaction-induced dispersive waves for trapping, we injected a control laser into a high-Q Si$_3$N$_4$ microcavity to enhance a comb line via injection locking or Kerr-induced synchronization  \cite{Srinivasan_Nature2023kerr,Yi_NP2025microcavity} and use it as an artificial dispersive wave for trapping (Fig. \ref{Fig1}). By phase modulating the injected laser, the soliton is observed to move synchronously with the modulated trapping well or dynamically trapped. Thus, soliton timing-engineering is achieved in an all-optical way, and we refer to the resulting microcomb as optically timing-engineered microcomb (OTEM). The all-optical approach simplifies the system and enables a slew rate reaching 31.3 ps/$\mu$s, more than 500 times faster than fiber-based TPFCs \cite{Newbury_Nature2022time,Newbury_NP2024free}.  A theory was derived to describe the fast response of the OTEM and adds insights to soliton dynamics subject to a modulated trapping well. 

\begin{figure*}[t!]
\begin{centering}
\includegraphics[width=0.98\linewidth]{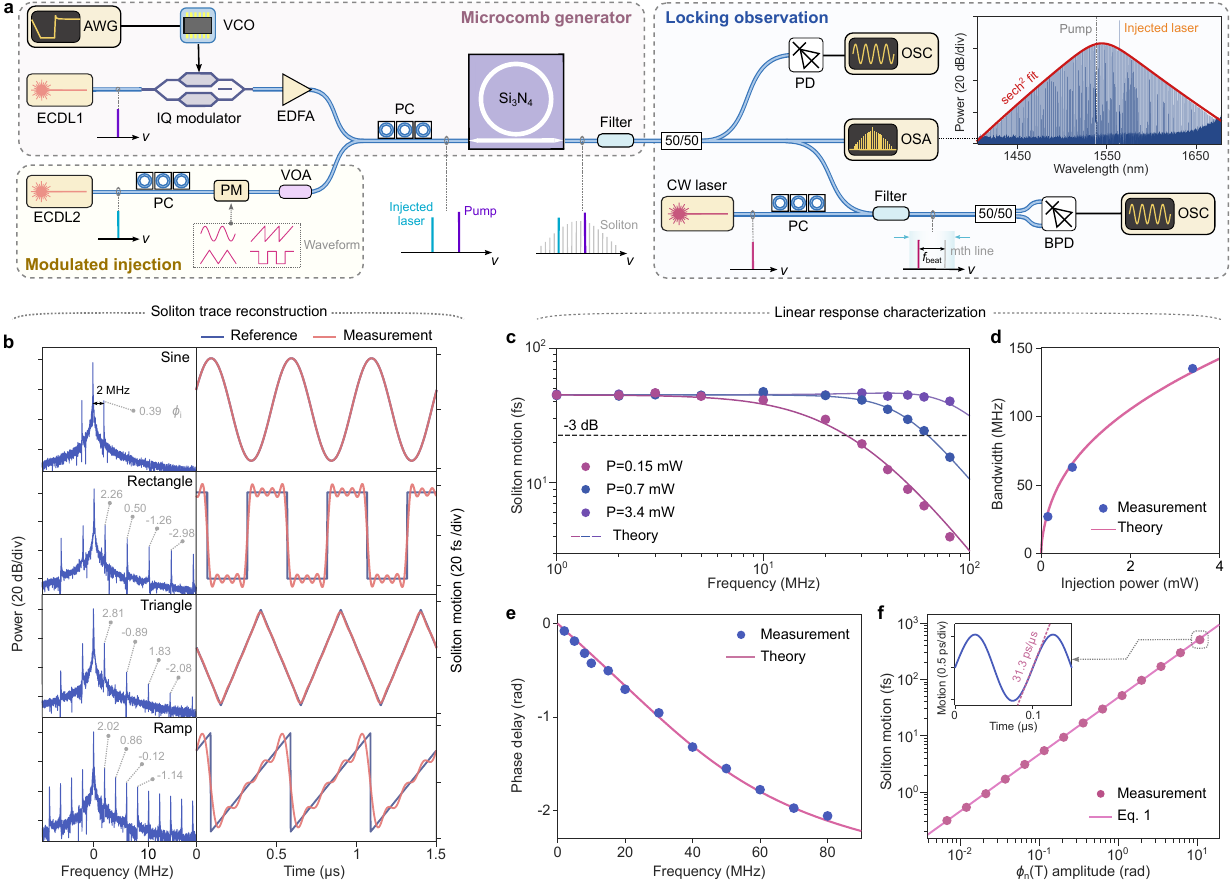}
\captionsetup{singlelinecheck=off, justification = RaggedRight}
\caption{\textbf{Timing-engineering of a soliton microcomb via dynamical trapping}.  \textbf{a,} Experimental setup to generate optically timing-engineered microcombs (OTEMs) and to characterize the soliton timing. The inset shows the spectrum of a soliton microcomb.
 \textbf{b,} Soliton motion retrieved by the heterodyne measurement under different drive waveforms. Numbers in the left panel show the phase of the corresponding sidebands. 
 \textbf{c,} Measured soliton motion amplitude under different sine-drive frequencies and injected optical powers. The measured response exhibits a bandwidth exceeding 100 MHz and fits the theory (Supplementary Sec. 1). 
\textbf{d,} Soliton response bandwidth increases with injected laser power and agrees with the theory.
\textbf{e,} Phase response of the OTEM under different modulation frequencies, in agreement with theory in Supplementary Sec. 1. 
 \textbf{f,} Retrieved soliton motion amplitude versus the amplitude of $\phi_n(T)$, in agreement with Eq. \ref{eq1}. The inset shows the retrieved soliton motion under $\phi_n^a$=10.7 rad. 
} 
\label{Fig2}
\end{centering}
\end{figure*}

\begin{figure*}[t!]
\begin{centering}
\includegraphics[width=0.98\linewidth]{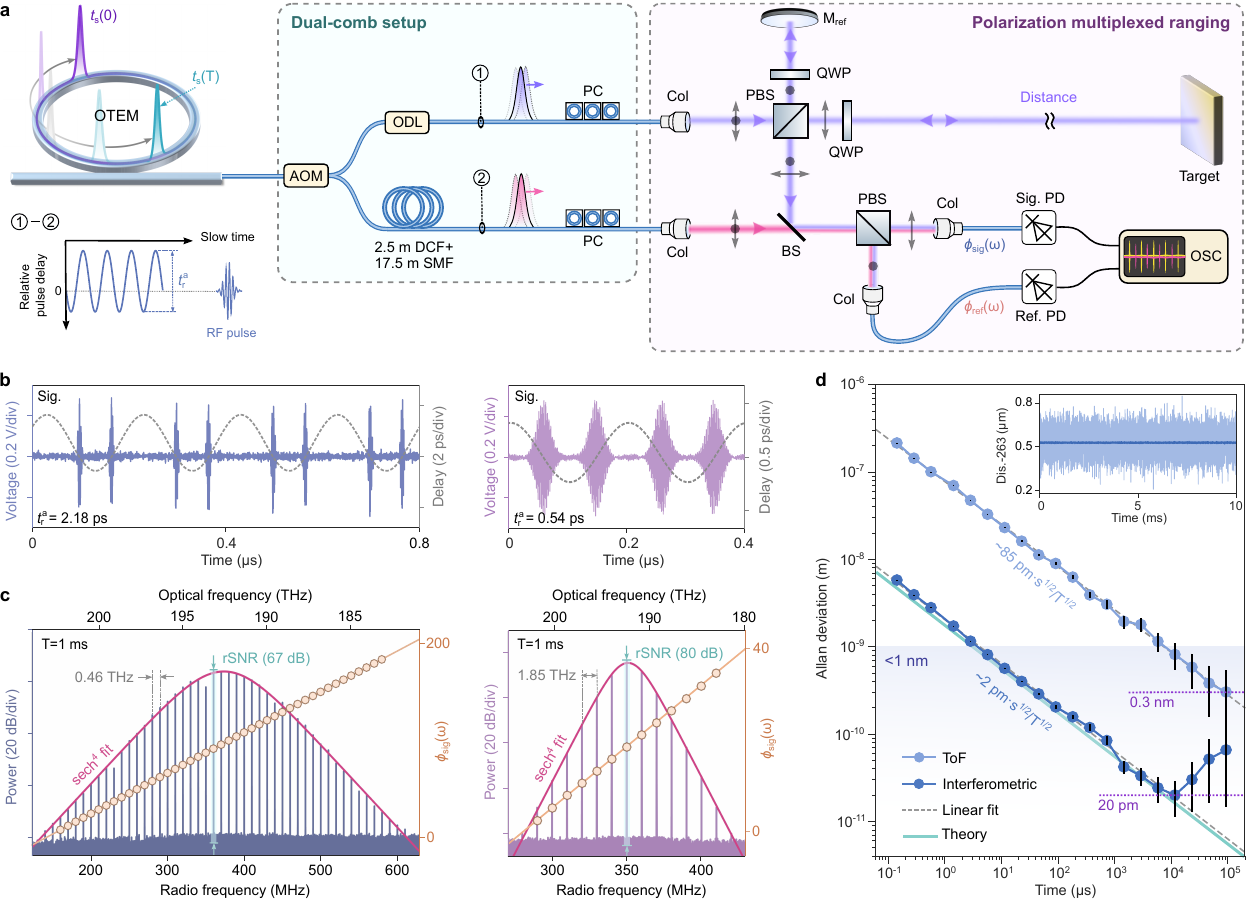}
\captionsetup{singlelinecheck=off, justification = RaggedRight}
\caption{\textbf{Single OTEM enabled dual-comb ranging.} \textbf{a,} Experimental setup for effective dual-comb ranging using an OTEM. Polarization-multiplexing was used to cancel the influence of fiber fluctuations in measurements. 
 \textbf{b,} Measured interferograms in the signal arm with different soliton scan span $t_r^a$. The dashed curves show the relative delay change between the split solitons. 
\textbf{c,} Optical power and phase spectra derived from the measured interferograms. The optical frequency resolution can be adjusted by $t_r^a$, and the spectral power envelopes have a sech$^4$-shape. The radiofrequency (RF) lines can reach a peak signal-to-noise ratio from the average noise (rSNR) of 67 or 80 dB within 1 ms. 
 \textbf{d,} Allan deviation of the absolute distance measured by the spectral phase fitted time-of-flight (ToF) and the interferometric phase of a single line, showing a normalized precision of 85 pm$\cdot$$\sqrt{\rm s}$ and 2 pm$\cdot$$\sqrt{\rm s}$, respectively. The inset shows the measured distance for these two methods.
}
\label{Fig3}
\end{centering}
\end{figure*}

\begin{figure*}[t!]
\begin{centering}
\includegraphics[width=0.9\linewidth]{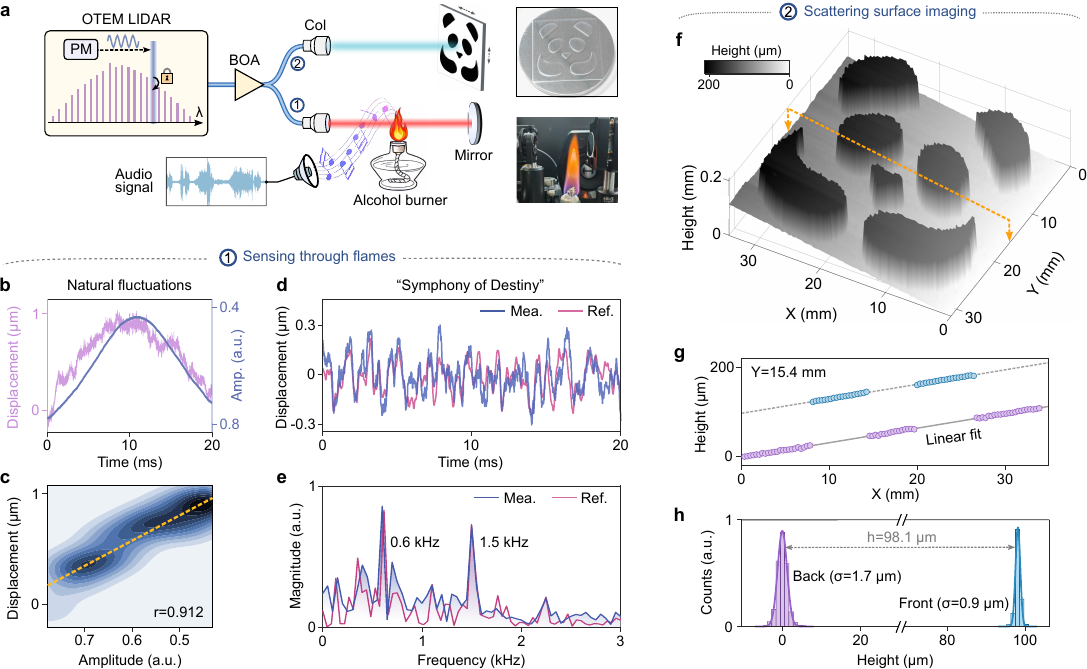}
\captionsetup{singlelinecheck=off, justification = RaggedRight}
\caption{\textbf{Absolute ranging and imaging using an OTEM.} 
\textbf{a,} Experimental setup for ranging through fluctuating flames and 3D imaging of a scattering aluminum plate whose reflection loss can exceed 50 dB.
\textbf{b,} Measured distance and optical power changes induced by a naturally fluctuating flame. 
\textbf{c,} Correlation between the distance and power changes. 
\textbf{d,} Measured distance change induced by a speaker playing the Symphony of Destiny, which agrees with the input signal. 
\textbf{e,} Audio spectrum of the measured distance change induced by the perturbed flames.
\textbf{f,} 3D imaging of the aluminum plate with a panda image. 
\textbf{g,} Distance change along the dashed slice in panel \textbf{f}. The solid and dashed lines are linear fits of the measured distance change for the front and back surfaces, respectively.
\textbf{h,} Residual error distribution for the measured distance subtracted by the solid linear fit. 
}
\label{Fig4}
\end{centering}
\end{figure*}

\begin{figure*}[t!]
\begin{centering}
\includegraphics[width=0.98\linewidth]{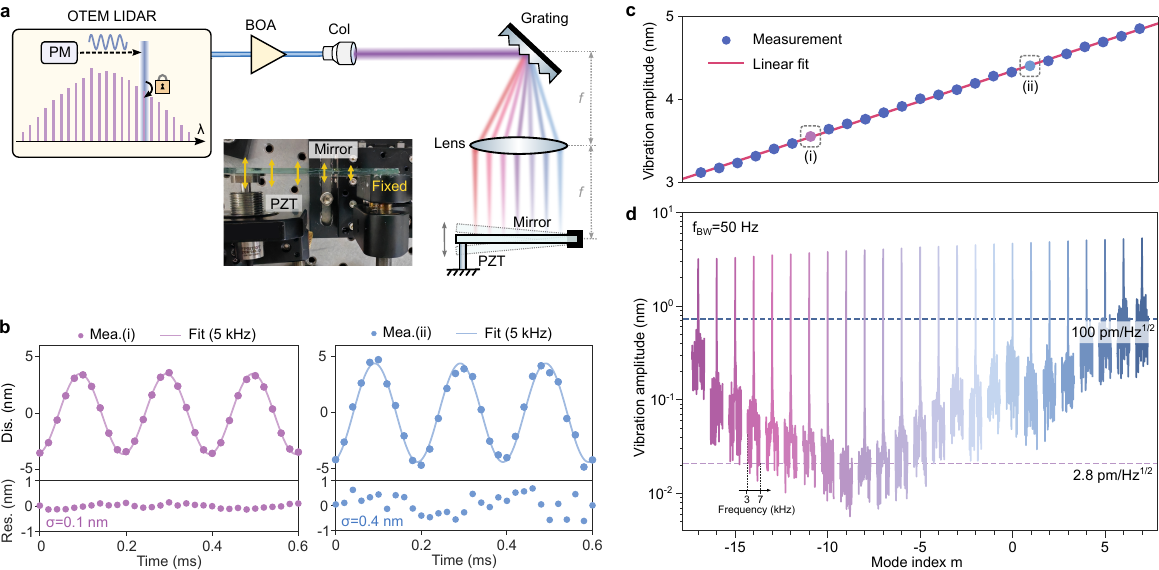}
\captionsetup{singlelinecheck=off, justification = RaggedRight}
\caption{\textbf{Parallel ranging using an OTEM.} 
\textbf{a,} Experimental setup for parallel ranging by dispersing the OTEM by a grating.
\textbf{b,} Measured displacement when the piezo drives the mirror at 5 kHz.
\textbf{c,} Measured displacement amplitude for different comb mode index, i.e., different positions of the mirror.
\textbf{d,} Spectral power density of vibrations for all the measured 25 comb modes with a resolution bandwidth of 50 Hz, reaching the highest sensitivity of 2.8 pm/$\sqrt{\rm Hz}$ (mode $-$9). 
}
\label{Fig4-2}
\end{centering}
\end{figure*}

Leveraging the dynamical trapping, we simplified dual-comb ranging (DCR) \cite{Newbury_NP2009rapid} by merely using a single timing-engineered comb instead of two combs. The OTEM effectively apodizes the interferogram, reducing photon waste in the dead zone of dual-comb measurements \cite{Newbury_Nature2022time}. The improved photon efficiency enabled the ranging precision to converge to 20 pm in 10 ms, marking a new record for optical absolute distance metrology \cite{Newbury_NP2009rapid,Bao_NC2025nanometric,Kim_NP2020ultrafast,Newbury_Nature2022time}, to our knowledge. 
The OTEM LIDAR was used  to sense optical-path changes through fluctuating flames and for 3D imaging. Spatially dispersing the OTEM further enables parallel one-dimension vibration measurement with a sensitivity as high as 2.8 pm/$\sqrt{\rm Hz}$ using a single photodetector.
The simplified OTEM architecture is fully compatible with foundry processed photonic integrated circuits (PICs) \cite{Liu_PR2023foundry} and can enable a versatile microcomb metrology platform with precision approaching the quantum limit \cite{Newbury_Nature2022time} (Fig. \ref{Fig1}).

\noindent \textbf{Principle of dynamical trapping and OTEM.} The working principle is illustrated in Fig. \ref{Fig1}. 
Intuitively, the trapping well formed by the injection-enhanced comb line moves within the microcavity, when the injected laser is modulated. For a trapped soliton, the soliton moves together with the trapping well. Thus, soliton timing ($t_s$) can be engineered via phase modulation. In the frequency domain, when locking the $n$th microcomb line by injection locking, two frequencies of the microcomb ($\nu_0$ for the pump and $\nu_n$ for the injected laser) are externally anchored. According to the Fourier transform theorem, phase slope between these two frequencies determines temporal position of the soliton as,

\begin{equation}
t_s(T)=\frac{\phi_n(T)-\phi_0}{2\pi nf_r},
\label{eq1}
\end{equation}
where $\phi_n(T)$ is the modulated phase of $\nu_n$, $\phi_0$ is the phase of the pump, and $T$ is a slow time. Frequency locking and timing-engineering are realized optically, which eliminates frequency locking circuits and enables fast response. A 2$\pi$ phase shift corresponds to a temporal delay of an optical cycle for a continuous wave (CW) laser, that is 5 fs at 1550 nm, but enables a temporal shift of 1/$nf_r$ for an OTEM, that can be a few picoseconds. Therefore, our approach acts as a magnifier in timing-engineering. Rigorous analysis of soliton timing $t_s$ in dynamical trapping using the Lagrangian approach \cite{Vahala_Optica2022efficiency,Matsko_OE2013Jitter} can be found in Supplementary Sec. 1. The analytical result (Supplementary Sec. 1) is consistent with the simplified Eq. \ref{eq1}.

Dynamically trapped solitons can enable coherent dual-comb metrology using a single OTEM. By splitting the OTEM into two arms and delay one of them by $T_m/2$ ($T_m$ is the modulation period of $\phi_n$), we attain a relative delay change $t_r(T)$=$t_s(T)-t_s(T+T_m/2)$ between the split solitons. This relative delay scan enables effective Fourier transform measurement or dual-comb measurement without any moving parts. 
Here, we used an OTEM for DCR to showcase the dynamical trapping-enabled dual-microcomb metrology. Its application in dual-comb spectroscopy \cite{Newbury_NP2024free,Newbury_Optica2016}, multi-dimensional coherent spectroscopy \cite{Cundiff_NP2018tri}, pump-probe spectroscopy \cite{Ye_Science2004united} and microwave photonics \cite{Capmany_NP2019integrated} can be anticipated (Fig. \ref{Fig1}). 

\noindent \textbf{Soliton timing-engineering.} We generate the soliton microcombs in a Si$_3$N$_4$ microresonator by fast laser frequency sweeping (Fig. \ref{Fig2}a, see Methods). To verify the above timing-engineering approach, we used our recently developed heterodyne beat measurement to retrieve the soliton motion trajectory \cite{Bao_PRX2025rhythmic}. By beating the $m$th microcomb line with a third CW laser, the amplitude and phase of motion harmonics can be retrieved (Eq. \ref{eqn3EXR} in Methods). 
We first phase modulated $\nu_n$ by a  sine-wave at 2 MHz. The heterodyne beat shows additional tones spaced by the modulation frequency $f_m$ from the main tone (Fig. \ref{Fig2}b). The reconstructed soliton motion trajectory agrees well with the theoretical trajectory derived using $\phi_n(T)$ and Eq. \ref{eq1}. We also drove the phase modulator by other waveforms at 2 MHz (e.g., triangle, square, and ramp waves). Our measurements retrieve all the soliton motion trajectories with high fidelity (Fig. \ref{Fig2}b). It is interesting to note that the spectra for square and triangle trajectories only exhibit odd motion harmonics. These measurements validate our timing-engineering approach via dynamical trapping.
 
Then, we characterized the response bandwidth by measuring the sine-motion amplitude $t_s^a$ under different drive frequencies with the sine-modulation amplitude (denoted as $\phi_n^a$) set at 0.91 rad. Figure \ref{Fig2}c shows the response bandwidth can exceed 100 MHz, much larger than fiber-based TPFCs that are limited by actuators and frequency locking circuits \cite{Newbury_Nature2022time}. This bandwidth can increase with the power of the injected laser in nearly a square-root way, as the theory predicts (Fig. \ref{Fig2}d and Supplementary Sec. 1).
Different from the simplified Eq. \ref{eq1}, Lagrangian analysis shows that there is a phase delay between the soliton motion $t_s(T)$ and modulation $\phi_n(T)$ for high frequency modulation (Supplementary Sec. 1). This phase delay was also measured to be in excellent agreement with our theory (Fig. \ref{Fig2}e and Supplementary Sec. 1). 

We also measured the motion amplitude $t_s^a$ under different $\phi_n^a$ with $f_m$ fixed at 10 MHz. $t_s^a$ increases with the phase modulation amplitude and follows Eq. \ref{eq1} (Fig. \ref{Fig2}f). For $\phi_n^a$=10.7 rad, the retrieved soliton motion trajectory is plotted in the inset of Fig. \ref{Fig2}f and the slew rate reached 31.3 ps/$\mu$s.

\noindent \textbf{Dual-comb metrology using OTEM.} 
The dynamical trapping enabled OTEM was used for dual-comb metrology with a simplified system. Following the concept in Fig. \ref{Fig1}, we built a DCR setup as shown in Fig. \ref{Fig3}a. A sine-wave, as opposed to a triangle-wave, was used for timing-engineering to avoid distortion from high harmonics due to a finite timing response bandwidth. In the experiment, we modulated the injected laser at an intermediate frequency of 5 MHz. Therefore, one of the arms was delayed by a $\sim$20 m fiber so as to have the required temporal delay of $T_m/2$=100 ns. To cancel fluctuations of the fibers, a polarization multiplexed setup was used for ranging \cite{lee2013absolute}. More details of the setup can be found in Methods. 

Examples of the measured interferograms with different timing scan spans $t_r^a$ are shown in Fig. \ref{Fig3}b. Due to the sine-modulation, relative delay $t_r(T)$ between the two split OTEM pulses changes sinusoidally (see right-axis of Fig. \ref{Fig3}b). This sinusoidal delay change was resampled into a linear delay change for spectral retrieval (Supplementary Sec. 2). Similar resampling was used in comb-calibrated frequency modulated continuous wave (FMCW) LIDAR using nonlinearly frequency swept lasers \cite{Newbury_OL2013comb,Bao_SA2025microcomb}. Fourier transform of the resampled data yields spectral power and phase shown in Fig. \ref{Fig3}c. The spectral envelope inherits a sech$^4$-shape from the sech$^2$-shaped soliton microcomb (RF power has a quadratic dependence on optical power), validating our approach. 
The line spacing in the RF domain is 2$f_m$=10 MHz, as a half of the OTEM scan cycle, either forward or backward scan, is cascaded for data processing. However, this RF line spacing corresponds to different spacings in the optical domain, which is determined as $f_{\rm res}=1/t_r^a$. The controllable resolution differs from conventional dual-comb systems \cite{Newbury_Optica2016}, where $f_{\rm res}$ always equals $f_r$. 

Since $t_r^a$ is smaller than $T_R$, the dead zone in conventional dual-comb measurements, where two comb pulses walk off, can be reduced. Therefore, the signal-to-noise ratio of the RF lines (rSNR), that is the RF line power versus the average noise floor, can be enhanced at expense of reduced spectral resolution. This enhancement can also be understood in the frequency domain as multiple optical comb lines contribute to a single RF line. In our experiment, the peak rSNR reaches 67 dB in 1 ms measurement time with $t_r^a$=2.18 ps, and this peak rSNR at 1 ms can be further boosted to 80 dB by reducing  $t_r^a$ to 0.54 ps (Fig. \ref{Fig3}c). Further measurements confirm that peak rSNR scales quadratically with 1/$t_r^a$ (Supplementary Fig. S3). Similar trend has also been observed in a fiber-laser-based dual-comb system by modulating the laser cavity length and switching the sign of repetition rate difference $\delta f_r$ \cite{Newbury_APLP2024broadband}. Here, we observe this trend by direct timing-engineering. 

The rSNR in Fig. \ref{Fig3}c is 21 or 34 dB higher than the recently demonstrated counter-propagating (CP) soliton-based DCR system \cite{Bao_NC2025nanometric} (over-coupling condition also contributes to this improvement, Methods). As the phase deviation of an RF line is determined as $\sigma_\phi=1/\sqrt{\rm rSNR}$ \cite{Bao_NC2025nanometric}, the enhanced rSNR can lead to higher ranging precision. For distance measurement, $t_r^a$ also determines the measurement range. Hence, we used a relatively large $t_r^a$=2.18 ps for ranging (corresponding to a slew rate of 18.2 ps/$\mu$s).

Then, we implemented DCR using the OTEM by measuring spectral phase in the signal arm $\phi_{\rm sig}(\omega)$ and the reference arm $\phi_{\rm ref}(\omega)$, see Fig. \ref{Fig3}a. A time-of-flight (ToF) can be derived by linearly fitting $\phi_{\rm sig}(\omega)-\phi_{\rm ref}(\omega)$ \cite{Newbury_NP2009rapid,Bao_NC2025nanometric}, and the derived absolute distance is shown in the inset of Fig. \ref{Fig3}d (Supplementary Sec. 4). As a half scan cycle is needed to retrieve the complex spectrum, the shortest measurement time is 100 ns. The normalized ranging precision reaches 85 pm$\cdot\sqrt{\rm s}$ (6-fold improvement compared to our recent CP soliton LIDAR \cite{Bao_NC2025nanometric}), and the absolute precision converges to 0.3 nm at 0.1 s. When this DCR precision converges below $\lambda_c$/4 ($\lambda_c$ is a wavelength), the interferometric phase of a single line can be used for ranging to further enhance precision \cite{Newbury_NP2009rapid} (Supplementary Sec. 4). Leveraging this interferometric phase, our system reaches a record normalized precision of 2 pm$\cdot\sqrt{\rm s}$, in agreement with theory (Eq. \ref{eqnPrecision}). The highest absolute precision reaches 20 pm at 10 ms and further improvement can be hindered by path fluctuations in the setup. 

\noindent \textbf{Absolute ranging and imaging.}
Our system is capable of measuring distance through fluctuating flames (path-1 in Fig. \ref{Fig4}a), 
which is important to industrial safety and combustion diagnostics \cite{mitchell_Optica2018coherent}. However, ranging or imaging through flames faces challenges from parasitic flame radiation, and signal attenuation by scattering from soot and refractive index gradients \cite{mitchell_Optica2018coherent,locatelli2013imaging}. Our system distinguishes intensity and delay change induced by flames. Specifically, the intensity change can be read out by the amplitude of the interferogram, while the spectral phase enables absolute distance sensing. Figure \ref{Fig4}b shows an example of the measured comb power and distance change induced by naturally fluctuating flames. The distance and intensity changes are closely, but not fully correlated (Fig. \ref{Fig4}c); high frequency distance fluctuations are absent in intensity changes (Fig. \ref{Fig4}b). Then, we used a speaker to perturb the flame by a passage of the Symphony of Destiny. The measured distance change is in close agreement with the speaker output (Fig. \ref{Fig4}d). The power spectrum of the measured distance change exhibit strong tones at 0.6 and 1.5 kHz (the symphony was played by 2X speedup). The slight discrepancy can be attributed to natural fluctuations of the flames.  

As an absolute ranging technique, the OTEM LIDAR also measures step-like distance change. Using the path-2 in Fig. \ref{Fig4}a, we measured the absolute distance using an aluminum plate with a panda image as the target, whose reflection loss can exceed 50 dB. By scanning the target, we mapped a 3D panda image shown in Fig. \ref{Fig4}f. Since we slightly tilted the plate, the measured 3D image has a distance change along the x-axis. The change along the dashed line in Fig. \ref{Fig4}f is plotted in the top panel of Fig. \ref{Fig4}g. The front and back surfaces of the panda image clearly show two linear lines. Abrupt distance changes with an amplitude about 100 $\mu$m were measured. We further subtract the solid linear fit from the measured distance in Fig. \ref{Fig4}g, and distribution of the residual error is plotted in Fig. \ref{Fig4}h. The error results from the surface roughness of the aluminum plate, rather than our LIDAR. An independent linearity measurement shows our system has a standard deviation of residual errors of 33 nm within the 360 $\mu$m measurement range (Supplementary Fig. S4). The mechanical scan of the target can be replaced by beam steering using an optical phase array for fast, precise 3D imaging in the future \cite{Watts_OL2017coherent}.

\noindent \textbf{Parallel ranging.} The broad bandwidth was further harnessed for parallel ranging by spatially dispersing the OTEM \cite{Vahala_Optica2019microresonator,Kippenberg_Nature2020massively}. Here, the OTEM was dispersed in one dimension via a grating, but two-dimensional dispersion is also possible \cite{Vahala_Optica2019microresonator}. The dispersed light was reflected from a mirror that was attached to a piezo at one of its edges (Fig. \ref{Fig4-2}a). When the piezo was driven, different frequencies experience distance changes with different amplitudes. Figure \ref{Fig4-2}b shows two examples of measured displacement at two positions of the mirror. Both recover the sine-trajectory with amplitudes of 3.55 nm and 4.41 nm, respectively. The residual errors are 0.1 nm and 0.4 nm with a single measurement time of 20 $\mu$s. A total of 25 RF comb lines were measured (corresponding to an optical bandwidth of 11 THz), whose sine-amplitudes are plotted in Fig. \ref{Fig4-2}c. The measured amplitudes exhibit an excellent linear relationship with the mode index, showcasing the parallel sensing capability of our OTEM.

With the measured displacement change in the time domain, we analyzed the power spectral density (PSD) of the measured vibration  (Fig. \ref{Fig4-2}d). Sharp peaks at 5 kHz were observed for all the 25 measured positions. The average noise floor of the PSD can be used as a measure of the sensitivity of the displacement measurement. The highest sensitivity reaches 2.8 pm/$\sqrt{\rm Hz}$ for mode index $-$9. Despite comb line power variation across the spectrum, the sensitivity is no worse than 100 pm/$\sqrt{\rm Hz}$ for all the measured lines. Limited by the response time of the piezo, we only drove it at a rate of 5 kHz. In principle, one-dimensional vibration rates up to $f_m$/2 can be measured by our OTEM, considering the Shannon-Nyquist sampling theorem. This may enable 2D measurements of fast vibrating samples such as optomechanical membrane \cite{Wilson_PRAppl2023membrane}.

\noindent \textbf{Discussions.} We have achieved agile steering of pulse timing for microcavity solitons via dynamical trapping. Injection locking enables fast modulation of the trapping well to steer solitons in a simplified, all-optical architecture. 
The resulting OTEM has pushed the precision and speed of optical absolute ranging to an unprecedented regime  (Supplementary Fig. S5). To our knowledge, prior absolute ranging reports (not limited to DCR) can hardly break the 0.1 nm barrier. The improved photon efficiency and stable interferometric phase in our DCR enables tens-of-picometer precision. Further improvement toward quantum-limited metrology can be possible by operating in a ``tracking mode" \cite{Newbury_Nature2022time}.

All the building blocks for OTEM-metrology can be integrated on the Si$_3$N$_4$ platform. For example, the pump and control laser can be heterogeneously integrated \cite{Bowers_Science2021laser}, while electro-optical or acousto-optical modulation can be realized by heterogeneously integrating with lithium niobate \cite{Kippenberg_Nature2023ultrafast}. Ultra-low loss Si$_3$N$_4$ spiral cavity or waveguide with a length of 14 m has been demonstrated \cite{LiJ_SA2024chip} and can be used as the delay line. Therefore, dynamical trapping and OTEM should enable monolithic dual-comb metrology chips. This can shift the paradigm of dual-comb metrology by using a single OTEM rather than two combs. OTEM can also be realized in other integrated microcombs including self-injection locked solitons \cite{Bowers_Nature2020integrated,Bowers_Science2021laser}, dark pulses \cite{Weiner_NP2015mode}, and Pockels solitons \cite{Tang_NP2021pockels} to further enhance the dual-comb metrology capability. In particular, the pump and the soliton locate in two distinct spectral ranges for Pockels solitons. How to arrange the anchor frequencies for efficient timing-engineering can shed further light on soliton physics. The approach to steer soliton by modulating the trapping well may also be used to control matter wave solitons in Bose–Einstein condensates \cite{becker2008oscillations}. 

\vspace{3 mm}

\noindent{\bf Methods}

{\small

\noindent \textbf{Soliton generation and injection locking.} The intrinsic and loaded Q-factors of the microresonator are 8.6$\times$10$^6$ and 2.9$\times$10$^6$, respectively. The microresonator was pumped at 1536.6 nm with an on-chip pump power of about 200 mW. To mitigate thermal instability, the soliton was generated by single sideband modulator based fast laser sweeping technique  \cite{Bao_PRX2025rhythmic}. Due to the over-coupling condition, the microcomb has a relatively high on-chip power of about 1.8 mW. To lock the $n$th line for timing-engineering, we tuned the injected laser to a frequency close to $\nu_n$ at 192.03 THz ($n$=$-$32). 
The on-chip power for the injected laser was typically adjusted between 1.5 and 2.5 mW. We observed a locking range of $\sim$75 MHz, when the injected laser was tuned around the $n$th line. 
For measurements in Fig. \ref{Fig3}, the phase modulator was driven at 5 MHz with an RF power of 33 dBm, while $V_\pi$ of the modulator is 3.7 V, yielding an amplitude of $\phi_n(T)$ as 12.2 rad. 

\vspace{1 mm}

\vspace{1 mm}
\noindent \textbf{Soliton motion trajectory retrieval.} Periodic soliton motion creates sidebands around the main comb mode. The complex amplitude extinction ratio between the main comb mode and the sidebands was derived as (see ref. \cite{Bao_PRX2025rhythmic} for details),

\begin{equation}
r(m,m_i) \approx -\frac{ie^{i\phi_{mi}}}{m\pi f_r \delta t_{mi}},  
\label{eqn3EXR}
\end{equation}
where $m$ is the comb mode number with respect to the pump, $\delta t_{mi}$ and $\phi_{mi}$ are the motion amplitude and phase in the $m_i$th harmonic, respectively. Thus, the motion trajectory can be retrieved by beating the $m$th line with a CW laser to measure $r(m,m_i)$, which gives $\delta t_{mi}$ and $\phi_{mi}$ to recover the motion trajectories shown in Fig. \ref{Fig2}b. We used $m=-$17 and $m_i \in [1,6]$ for reconstruction.

\vspace{1mm}
\noindent \textbf{Ranging setup.}
The fiber link in Fig. \ref{Fig3}a includes a piece of 2.5 m dispersion compensation fiber (DCF) to manage the net dispersion and minimize temporal broadening of the soliton pulses. Since the peak-to-peak temporal scan span $t_r^{a}$ in our experiment is smaller than a round-trip time $T_R$, we finely tuned the relative delay within $T_R$ to overlap the solitons temporally for effective dual-comb measurements (see optical delay line, ODL, in Fig. \ref{Fig3}a). We also frequency shifted one of the arms by 350 MHz using an acousto-optical modulator (AOM). This shift avoids spectral aliasing and is critical to data processing. 
For the measurements in Fig. \ref{Fig3}, the received signal comb power and local comb power were $\sim$10 $\mu$W and $\sim$30 $\mu$W, respectively, considering the loss in the ranging setup. No optical amplification was used for the measurement in Fig. \ref{Fig3}. 

For the measurements in Figs. \ref{Fig4}, \ref{Fig4-2}, the microcomb was amplified by a semiconductor optical amplifier (Thorlabs BOA1550S) to  compensate the higher loss in the setup. The BOA-amplifier has a broad amplification bandwidth of 105 nm around 1550 nm, which enables amplification of more than 120 microcomb lines. The microcomb was amplified to about 3 mW for measurements in Figs. \ref{Fig4} and 10 mW for measurements in Figs. \ref{Fig4-2}. The received signal comb powers were 20--65 $\mu$W for Figs. \ref{Fig4}b-e, 1 nW--350 $\mu$W for Figs. \ref{Fig4}f-h and $\sim$50 $\mu$W for Figs. \ref{Fig4-2}, respectively.

\vspace{1mm}
\noindent \textbf{Theoretical ranging precision.} Since the phase deviation is determined as 1/$\sqrt{\rm rSNR}$ \cite{Bao_NC2025nanometric}, the ranging precision using the interferometric phase can be derived as,

\begin{equation}
\sigma_{\rm L}(T)=\frac{\sqrt{(1+\alpha)}\lambda_c\sigma_{\phi}(T)}{4\pi}=\frac{\sqrt{(1+\alpha)}\lambda_c}{4\pi\sqrt{\text{rSNR}(m,T)}},  
\label{eqnPrecision}
\end{equation}
where $\alpha$ is rSNR ratio between the signal and reference arms ($\alpha$=0.87) and $\lambda_c$=1543.974 nm is the used carrier wavelength, $m$ is the used mode index.  The theoretical precision is plotted in Fig. \ref{Fig3}d.






\vspace{3 mm}
\noindent \textbf{Data Availability.}
The data that support the plots within this paper and other findings are available in the paper.

\noindent \textbf{Acknowledgements.}
We thank Prof. Qiang Liu and Prof. Yidong Tan at Tsinghua University for discussions and equipment loan. The silicon nitride chip used in this work was fabricated by Qaleido Photonics. This work is supported by the Beijing Natural Science Foundation (JQ25016), by the National Key R\&D Program of China (2023YFB3211200, 2021YFB2801200), by the National Natural Science Foundation of China (62250071, 62175127, 62375150), and by the Tsinghua-Toyota Joint Research Fund. J.L. acknowledges support from the National Natural Science Foundation of China (12261131503), Innovation Program for Quantum Science and Technology (2023ZD0301500), Shenzhen-Hong Kong Cooperation Zone for Technology and Innovation (HZQB-KCZYB2020050), and Shenzhen Science and Technology Program (RCJC20231211090042078).

\vspace{1 mm}

\noindent\textbf{Author Contributions.} Z.W. led the experiments and analysis with assistance from G.L., C.L., and C.Y.; Y.H., Z.Z. and J.L. prepared and characterized the Si$_3$N$_4$ chip. The project was supervised by C.B.

\vspace{1 mm}
\noindent \textbf{Competing Interests.} The authors declare no competing interests.

\bibliography{main}


\end{document}